\documentclass[prd,twocolumn,floatfix,amsmath,nofootinbib,amssymb,preprintnumbers,showkeys]{revtex4}

\usepackage{graphicx,color,dcolumn,booktabs,bm,multirow}
\usepackage{longtable,lscape}
\usepackage{txfonts}
\usepackage{overpic}
\usepackage{array}
\usepackage{indentfirst}
\usepackage{feynmf}   %{feynmp}
\usepackage{slashed}  %for Feynman symbols
\usepackage{cases}
\usepackage{epstopdf}
\usepackage{tabularx}
\usepackage{braket}
\usepackage{CJKutf8}
\usepackage{footnote}
\usepackage{physics}
\usepackage{accents}
\usepackage[colorlinks,
citecolor=blue,
anchorcolor=red,
menucolor=red,
linkcolor=red,
filecolor=red,
runcolor=red,
urlcolor=blue,
frenchlinks=red]{hyperref}

\begin{document}
\begin{CJK}{UTF8}{gbsn}

\title{$T_{\bar{c}\bar{s}1}^f(2750)$ production in the $B^+$ decays processes}
\author{Zhuo Yu$^{1}$}\email{zhuo@seu.edu.cn}
\author{Qi Wu$^{2}$}\email{wuqi@htu.edu.cn}
\author{Zi-Li Yue$^{1}$}t\email{yuezili@seu.edu.cn}
\author{Dian-Yong Chen$^{1,3}$\footnote{Corresponding author}}\email{chendy@seu.edu.cn}
\affiliation{$^1$ School of Physics, Southeast University, Nanjing 210094, China}
\affiliation{$^2$Institute of Particle and Nuclear Physics, Henan Normal University, Xinxiang 453007, China}
\affiliation{$^3$ Lanzhou Center for Theoretical Physics, Lanzhou University, Lanzhou 730000, China}
\date{\today}

\begin{abstract}
In the present work, we studied the $T_{\bar{c}\bar{s}1}^f(2750)$ state production through the meson loop mechanism in the $B^+$ meson decays, where $T_{\bar{c}\bar{s}1}^f(2750)$ is considered as a $\bar{D}K^*$ molecular state with $J^P=1^+$. By employing the effective Lagrangian approach, we estimated the branching ratio of the $B^+ \to D^+ T_{\bar{c}\bar{s}1}^ff(2750) $ and $B^+ \to D^{*+} T_{\bar{c}\bar{s}1}^f(2750)$ processes and found them to be on the order of $10^{-5}\sim 10^{-4}$. The fit fraction of $T_{\bar{c}\bar{s}1}^f(2750)$ in different processes was also estimated. We propose to search for $T_{\bar{c}\bar{s}1}^f(2750)$ in the $B^+ \to D^{*+}\bar{D}^{*-}K^+$ process, which should be accessible to the Belle II and LHCb Collaborations.
\end{abstract}
\maketitle

\section{introduction}
\label{sec:introduction}
In the past two decades, an increasing number of tetraquark and pentaquark candidates have been observed experimentally. In the hidden-charm sector, the charged $Z_c(3900)$ observed by the BESIII and Belle Collaborations~\cite{BESIII:2013ris,Belle:2013yex} and the $P_c(4312/4440/4457)$ states observed by the LHCb Collaboration~\cite{LHCb:2019kea} have been widely accepted as candidates for hidden charm tetraquark and pentaquark states, respectively. In the open charm sector, significant progress has also been made in recent years in the study of charmed tetraquark states containing a single charm quark.

In 2003, the BaBar Collaboration reported the observation of $D_{s0}^*(2317)$ in the $D_s^+\pi^0$ invariant mass spectrum from $e^+e^-$ annihilation process, with a mass near $2.32$ GeV~\cite{BaBar:2003oey}, which was subsequently confirmed by both the CLEO and Belle Collaborations~\cite{CLEO:2003ggt,Belle:2003guh}. In addition to $D_{s0}^*(2317)$, the CLEO Collaboration observed another structure, $D_{s1}(2460)$, with a mass around $2.46$ GeV in the $D_s^{*+}\pi^0$ invariant mass distribution, which was later confirmed by other experiments~\cite{Belle:2003guh,BaBar:2003cdx,Belle:2003kup}.

The $J^P$ quantum numbers are consistent with those of the $P$ wave charm-strange mesons, which indicates that these two states could be $P-$wave $c\bar{s}$ mesons with $J^P=0^+$ for $D_{s0}^*(2317)$ and $J^P=1^+$ for $D_{s1}(2460)$~\cite{Colangelo:2003vg,Wei:2005ag,Wang:2006fg,Wang:2006zw,Matsuki:2011xp,Ke:2013zs}. However, their measured masses are significantly lower than the predictions of the conventional quark model~\cite{Godfrey:1985xj,Godfrey:2003kg}. This discrepancy has motivated several alternative interpretations. Given that the masses of these two states lie close to the $DK$ and $D^*K$ thresholds, it is natural to assign them as $DK$ and $D^*K$ molecules, respectively~\cite{Guo:2006fu,Faessler:2007gv,Faessler:2007us,Cleven:2014oka,Wu:2019vsy,Zhu:2019vnr,Liu:2022dmm,Yue:2023qgx,Liu:2023cwk,Kim:2023htt, Yue:2025wcl, Xiao:2016hoa}. Other interpretations, such as compact tetraquark states~\cite{Terasaki:2003qa,Dmitrasinovic:2004cu,Maiani:2004vq,Chen:2016spr,Maiani:2024quj} and $P-$ wave $\bar{c}s$ core couplings to $D^{(*)}K$ components which can dynamically generate the $D_{s0}^*(2317)$ and $D_{s1}(2460)$~\cite{vanBeveren:2003kd,Simonov:2004ar} have also been proposed.

Upon the observations of the $D_{s0}^*(2317)$ and $D_{s1}(2460)$, it is natural to search for other possible states near the $D^{(*)}K^{(*)}/\bar{D}^{(*)}K^{(*)}$ threshold. It should be mentioned that, with the exception of the $D^{*0}K^+$ and $D^{*+}K^0$ channels, states in other channels cannot be conventional meson resonances and thus must be open-charm tetraquark states. In particular, resonances in the $D^{\pm}K^+$ and $D^0K^0/\bar{D}^0 K^0$ channels would manifestly be exotic, having minimal quark content with four different flavors. Over the past five years, experimental searches for charmed tetraquark states near the $D^{(*)}K^{(*)}/\bar{D}^{(*)}K^{(*)}$ thresholds have yielded significant progress, as summarized in in Table.~\ref{tab:exp.information}. In the following, we briefly review these developments from both experimental and theoretical perspectives.

\begin{table*}[htbp]
\centering
\caption{Experimental measurements of charmed tetraquark states.}
\label{tab:exp.information}
\begin{ruledtabular}
\begin{tabular}{lllllll}
%\hline
State & Channel & Mass (MeV) & Width (MeV) & $I(J^P)$ & Threshold & Experiment \\
\hline
$D_{s0}^{*}(2317)$ & $D_{s0}^{*}(2317)\to D_s^+ \pi^0$ & $\sim2320$ & $-$ & $0(0^+)$ & $DK$ & BABAR~\cite{BaBar:2003oey} \\[2ex]
                   & $D_{s0}^{*}(2317)\to D_s^{+} \pi^0$ & $\sim2320$ & $-$ & $-$                   &  & CLEO~\cite{CLEO:2003ggt} \\[2ex]
                   & $B\to\bar{D}D_s\pi$ & $-$ & $-$ & $-$               &  & Belle~\cite{Belle:2003guh} \\[2ex]
$D_{s1}(2460)$ & $D_{s1}(2460)\to D_s^{*+} \pi^0$ & $\sim 2460$ & $< 3.8$ & $0(1^+)$ & $D^* K$ & CLEO~\cite{CLEO:2003ggt} \\[2ex]
$T_{\bar{c}\bar{s}0}^{*}(2870)^{0}$ & $B\to D^+ D^- K^+$ & $2866 \pm 7 \pm 2$ & $57 \pm 12 \pm 4$ & $0(0^+)$ & $\bar{D}^* K^*$ & LHCb~\cite{LHCb:2020bls,LHCb:2020pxc} \\[2ex]
                                    & $B^+\to D^{*+} D^- K^+$ & $2914 \pm 11 \pm 15$ & $128 \pm 22 \pm 23$ &  &  & LHCb~\cite{LHCb:2024vfz} \\[2ex]
                                    & $B^-\to D^- D^0 K^0_S$ & $2883 \pm 11 \pm 8$ & $87^{+22}_{-47}\pm17$ &  &                 & LHCb~\cite{LHCb:2024xyx} \\[2ex]
$T_{\bar{c}\bar{s}1}^{*}(2900)^{0}$ & $B\to D^+ D^- K^+$ & $2904 \pm 5 \pm 1$ & $110 \pm 11 \pm 4$ & $0(1^+)$ &  $\bar{D}^* K^*$  & LHCb~\cite{LHCb:2020bls,LHCb:2020pxc} \\[2ex]
                                    & $B^+\to D^{*+} D^- K^+$ & $2887 \pm 8 \pm 6$ & $92 \pm 16 \pm 16$ &  &  & LHCb~\cite{LHCb:2024vfz} \\[2ex]
$T_{c\bar{s}0}^{*}(2900)^{0}$ & $B^0\to \bar{D}^0 D^+_s \pi^-$ & $2892 \pm 14 \pm 15$ & $119 \pm 26 \pm 13$ & $1(0^+)$ & $D^* K^*$ & LHCb~\cite{LHCb:2022sfr,LHCb:2022lzp} \\[2ex]
$T_{c\bar{s}0}^{*}(2900)^{++}$ & $B^+\to D^- D^+_s \pi^+$ & $2921 \pm 17 \pm 20$ & $137 \pm 32 \pm 17$ & $1(0^+)$ & $D^* K^*$ & LHCb~\cite{LHCb:2022sfr,LHCb:2022lzp} \\[2ex]
$T_{c\bar{s}0}(2327)$ & $D_{s1}(2460)^+\to D_s^+ \pi^+ \pi^-$ & $2327 \pm 13 \pm 13$ & $96 \pm 16 \,^{+170}_{-23}$ & $1(0^+)$ & $DK$ & LHCb~\cite{LHCb:2024iuo} \\
%\hline
\end{tabular}
\end{ruledtabular}
\end{table*}

In 2020, the LHCb Collaboration announced the observation of two states, denoted as $X_0(2900)$ and $X_1(2900)$ (now referred to  $T_{\bar{c}\bar{s}0}^*(2870)$ and $T_{\bar{c}\bar{s}1}^*(2900)$, respectively~\cite{ParticleDataGroup:2024cfk}), in the $D^-K^+$ invariant mass distributions of the decay process $B^+ \to D^+D^-K^+$\cite{LHCb:2020bls,LHCb:2020pxc}. This is the first observation of manifestly exotic hadrons with a single charm quark. The analysis indicated that the $J^P$ quantum number of $T_{\bar{c}\bar{s}0}^*(2870)$ and $T_{\bar{c}\bar{s}1}^*(2900)$ are $0^+$ and $1^-$, respectively. Recently, the LHCb Collaboration confirmed the existence of both $T_{\bar{c}\bar{s}0}^*(2870)$ and $T_{\bar{c}\bar{s}1}^*(2900)$ in the $B^+\to D^{*+} D^- K^+$ decay~\cite{LHCb:2024vfz}, and observed $T_{\bar{c}\bar{s}0}^*(2870)^*$ in the $B^-\to D^- D^0 K^0_S$ decay without finding evidence for $T_{\bar{c}\bar{s}1}^*(2900)$~\cite{LHCb:2024xyx}. In 2022, the first doubly charged tetraquark candidate, $T_{c\bar{s}0}^a(2900)^{++}$, alone with its neutral isospin partner $T_{c\bar{s}0}^a(2900)^{0}$, were observed by the LHCb Collaboration in the $D_s \pi$ mass distribution of the $B^0 \to \bar{D}^0 D_s^+ \pi^-$ and $B^+ \to D^- D_s^+ \pi^+$ decays\cite{LHCb:2022sfr,LHCb:2022lzp}. The superscript $a$ here indicates the isospin is equal to 1. Very recently, the LHCb Collaboration reported $T_{c\bar{s}0}(2327)^{++}$ and its isospin partner through an amplitude analysis of the $D_{s1}(2460)^+ \to D_s^+\pi^+\pi^-$ reaction~\cite{LHCb:2024iuo}.

The abundant experimental information on open-charm tetraquark states has prompted various theoretical explanations, such as the hadronic molecular states and compact tetraquark states, or they may also be caused by the triangle singularities. Given that the observed open-charm tetraquark states lie near the $D^{(*)}K^{(*)}/\bar{D}^{(*)}K^{(*)}$ threshold, it is natural to interpret them as hadronic molecular states composed of a charmed meson and a Kaon meson~\cite{Ke:2022ocs,Kong:2021ohg,Wang:2021lwy,Chen:2021erj,Agaev:2020nrc,Chen:2020aos,Mutuk:2020igv,Liu:2020nil,Xiao:2020ltm,Huang:2020ptc,Yue:2022mnf,
Chen:2022svh,Agaev:2022eyk,Duan:2023lcj,Wang:2023hpp,Huang:2023fvj,Wang:2024fsz}. However, other hadronic molecular configurations are also possible and deserve to be discussed. For instance, whether $T_{\bar{c}\bar{s}1}^*(2900)^*$ can be interpreted as a $\bar{D}_1 K$ molecular state has been investigated in Refs.~\cite{He:2020btl,Qi:2021iyv,Chen:2021tad} since $T_{\bar{c}\bar{s}1}^*(2900)^*$ lies just below the $\bar{D}_1 K$ threshold. As the open-charm tetraquark states are observed in the invariant mass spectra of $D^-K^+$ or $D^{(*)}_s\pi$, compact tetraquark interpretations have been proposed in Refs.~\cite{Karliner:2020vsi,He:2020jna,Cheng:2020nho,Guo:2021mja,Zhang:2020oze,Wang:2020xyc,Agaev:2021knl,Ozdem:2022ydv}. Besides, $T_{\bar{c}\bar{s}0}^*(2870)$ and $T_{\bar{c}\bar{s}1}^*(2900)$ can also be explained by triangle singularity~\cite{Liu:2020orv}. 

%under the $D^{(*)}K^{(*)}/\bar{D}^{(*)}K^{(*)}$ molecular state scenario,

\begin{figure}[htb]
	\centering
	\includegraphics[scale=0.4]{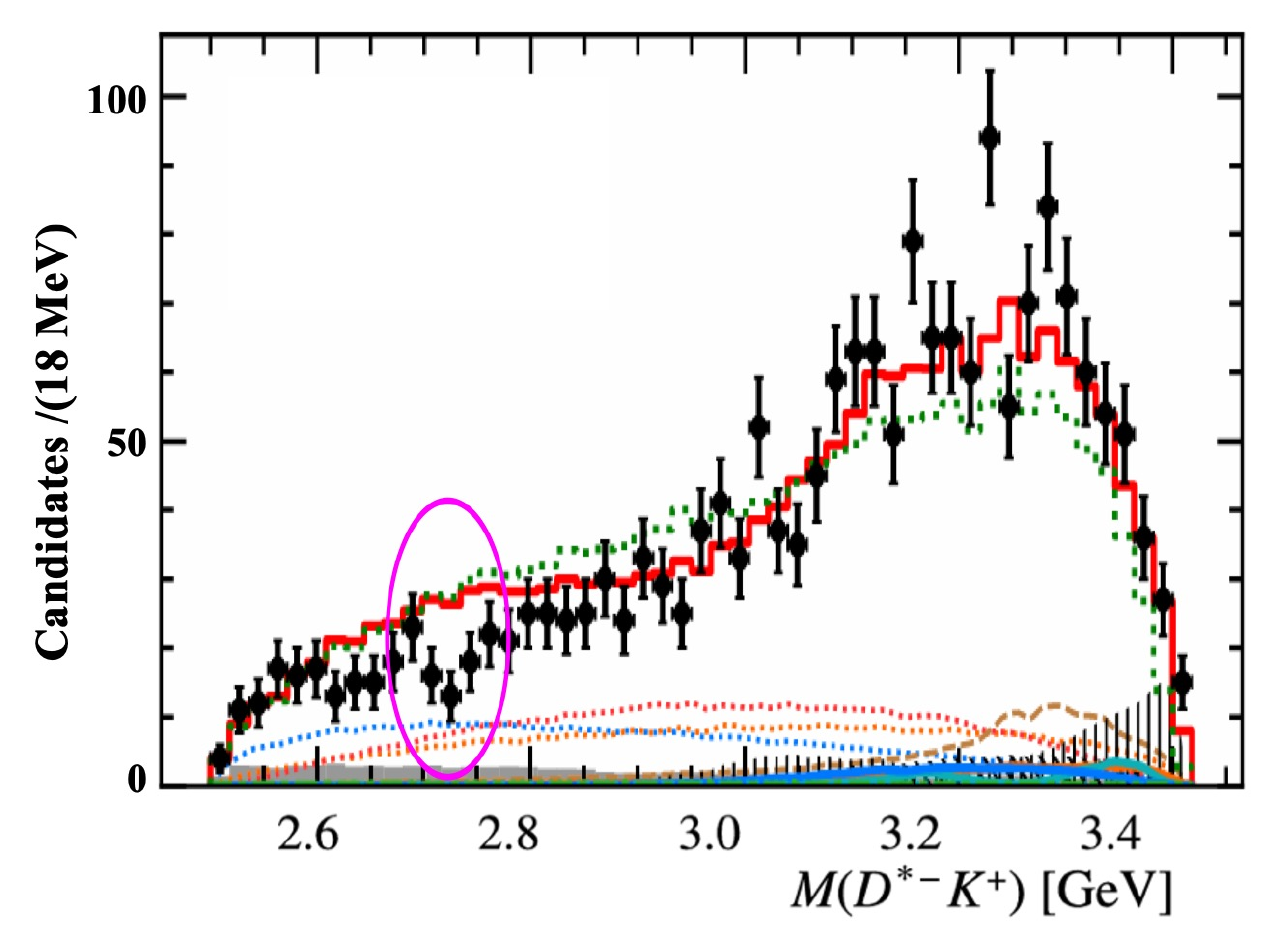}%temporary
	\caption{ Distributions of two-body invariant masses of $M(D^{*-}K^+)$ in the $B^+ \to D^{*-}D^+K^+$ reported by LHCb Collaboration (Fig.~1-(e) in Ref.~\cite{LHCb:2024vfz}).} \label{Fig:LHCb-D*K}
\end{figure}

To date, the $D_{s0}^*(2317)$ with $I=0$ and $T_{c\bar{s}}(2327)$ with $I=1$ in the $DK$ channel, the $D_{s1}(2460)$ in the $D^*K$ channel, and the $T_{\bar{c}\bar{s}0}^*(2870)^0$, $T_{\bar{c}\bar{s}1}^*(2900)^0$, and $T_{c\bar{s}0}^*(2900)^0$ in the $DK^*/\bar{D}K^*$ channel have been observed experimentally. However, although the $DK^*/\bar{D}K^*$ molecular states have been predicted by several  works\cite{Hu:2020mxp,Wang:2023hpp,Kong:2021ohg,Yang:2023wgu,Chen:2023syh}, they have not yet been experimentally confirmed. Actually, in the study of $B^+ \to D^{*+}D^-K^+$ and $B^+ \to D^{*-}D^+K^+$ decays by the LHCb Collaboration\cite{LHCb:2024vfz}, they mentioned that a structure around 2.7 GeV in the $D^{*-}K^+$ invariant mass distribution cannot be determined as an additional resonance or a mismodeling of one or more of the non-resonant components. For the convenience of readers, we present the $D^{*-}K^+$ invariant mass distributions in Fig.\ref{Fig:LHCb-D*K}, where one can find that a significant structure is observed around 2.7 GeV. In this way, we refer to this structure as the $S-$wave $\bar{D}K^*$ molecular state with $I=0$ as $T_{\bar{c}\bar{s}1}^f(2750)$ in the present work, and investigate its production properties in the $B^+$ meson decay process.

\begin{figure}[t]
	\begin{tabular}{cccc}
		\centering
		\includegraphics[width=4.2cm]{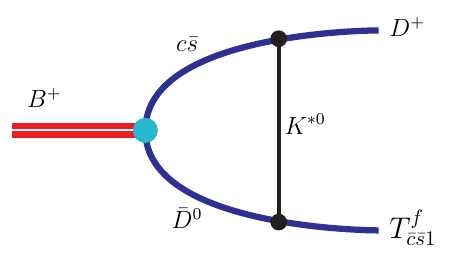}&
		\includegraphics[width=4.2cm]{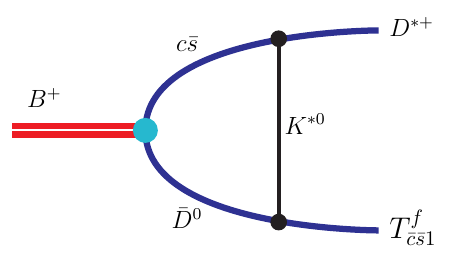}&\\
		\\
		$(a)$  & $(b)$ \\
	\end{tabular}
	\caption{Diagrams contributing to $B^+ \to D^{+} T_{\bar{c}\bar{s}1}^f(2750) $ (diagram (a)) and $B^+ \to D^{*+} T_{\bar{c}\bar{s}1}^f(2750) $ (diagram (b)) at the hadron level. Here, the c$\bar{s}$ refers to the considered charm-strange mesons.}\label{fig:hadronlevel}
\end{figure}

In our previous work, we have studied the productions of $X_0(2900)$ in the $\bar{D}^*K^*$ molecular framework in $B^+ \to D^{(*)+}X_0(2900)$~\cite{Yu:2023avh}. By considering the dominant contributions from triangle loops involving $D_{s1}^{\prime+}/D_{s0}^{*+} \bar{D}^{*0} K^{*0}$, we provided a good description of the production properties of $X_0(2900)$ in the molecular framework. Following this work, we investigate the $T_{\bar{c}\bar{s}1}^f(2750)$ production in $B^+ \to D^{(*)+}T_{\bar{c}\bar{s}1}^f(2750)$, where the anti-charmed meson $\bar{D}^{*0}$ in the loop is replaced by $\bar{D}^{0}$ since $T_{\bar{c}\bar{s}1}^f(2750)$ is assumed to be a $\bar{D}K^*$ molecular state. Furthermore, we include more charm-strange meson in the loop, mostly considering the sizable branching ratio of $B^+ \to (c\bar{s})^+ \bar{D}^0$ and the couplings of $(c\bar{s})^+ D^{(*)+} K^{*0}$. We present the branching ratios of $B^+ \to (c\bar{s})^+ \bar{D}^0$ in Table.~\ref{tab:bb}, and the relevant Feynman diagrams at the hadron level are shown in Fig.~\ref{fig:hadronlevel}. We employe the effective Lagrangian approach to calculate the triangle loop diagrams. In the following sections, we present the relevant effective Lagrangians and the details of how we calculate the production processes.

\begin{table}[t]
	\caption{The branching ratios of $B^+ \to (c\bar{s})^+ \bar{D}^0$ considered in the present work in PDG average~\cite{ParticleDataGroup:2022pth}\label{tab:bb}}
	\renewcommand\arraystretch{1.5}
	\begin{tabular}{p{3.5cm}<\centering p{4.5cm}<\centering }
		\toprule[1pt]
		%\midrule[0.65pt]
		Process & branching ratio \\
		\midrule[1pt]
		$B^+ \to D_s^+\bar{D}^0$ & $(9.3\pm0.6)\times10^{-3}$   \\
		$B^+ \to D_s^{*+}\bar{D}^0$ & $(7.6\pm1.6)\times10^{-3}$   \\
		$B^+ \to D_{s0}^{*+}\bar{D}^0$ & $>(7.9^{+1.5}_{-1.3})\times10^{-4}$   \\
		$B^+ \to D_{s1}^{\prime+}\bar{D}^0$ & $(3.1^{+1.0}_{-0.9})\times10^{-3}$   \\
		$B^+ \to D_{s1}^{+}\bar{D}^0$ & $>(4.0\pm1.0)\times10^{-4}$   \\
		%\midrule[0.65pt]
		\bottomrule[1pt]	
	\end{tabular}
\end{table}

%\begin{figure}[t]
%	\begin{tabular}{cc}
%		\centering
%		\includegraphics[width=6cm]{BtoDTcbarsbar1.pdf}& \\
%		(a) &\vspace{0.2cm} \\
%		\includegraphics[width=6cm]{BtoDxTcbarsbar1.pdf}&\\
%		(b)&\\
%	\end{tabular}
%	\caption{Diagrams contributing to $B^+ \to Z_{cs}^+ s\bar{s}/u\bar{u}$ at the quark level.}\label{Fig:quarklevel}
%\end{figure}

The rest of this work is organized as follows. After the introduction, we present the model used in the estimation of $T_{\bar{c}\bar{s}1}^f(2750)$ productions. The numerical results and relevant discussions  are given in Section.~\ref{Sec:Num}, and Section.~\ref{Sec:sum} is devoted to a brief summary.

\section{Theoretical framework}
\label{sec:theory}
\subsection{Effective Lagrangian}
In the present work, $T_{\bar{c}\bar{s}1}^f(2750)$ is assumed to be a $S$-wave bound state of $\bar{D}K^*$ with $I(J^P)=0(1^+)$, which is,
\begin{eqnarray}
	T_{\bar{c}\bar{s}1}^f=\frac{1}{\sqrt{2}}\left(|\bar{D}^0 K^{*0}\rangle+|D^- K^{*+}\rangle\right),
\end{eqnarray}
and the effective Lagrangian depicting the interaction of $T_{\bar{c}\bar{s}1}^f(2750)$ with its components can be,
\begin{eqnarray}
	\mathcal{L}_{T_{\bar{c}\bar{s}1}^f}=g_{T_{\bar{c}\bar{s}1}^f}T_{\bar{c}\bar{s}1}^{f\mu}\left(\bar{D}^0 K^{*0}_\mu + D^- K^{*+}_\mu \right),
\end{eqnarray}
where $g_{T_{\bar{c}\bar{s}1}^f}$ is the effective coupling constant.

As for the $B^+$ weak decay vertex,  we utilized the parametrized hadronic matrix elements obtained by the effective Hamiltonian at the quark level, which are~\cite{Cheng:2003sm,Soni:2021fky},
\begin{eqnarray}
\label{eq:weak transition}
&&\left\langle0|J_\mu|P(p_1)\right\rangle = -if_{P} p_{1\mu}\;,\nonumber\\
&&\left\langle0|J_\mu|V(p_1,\epsilon)\right\rangle =f_V \epsilon_\mu m_V \;,\nonumber\\
&&\left\langle P(p_2)|J_\mu|B(p_0)\right\rangle \nonumber\\&&\quad=\left[P_\mu-\frac{m^2_{B}-m^2_P}{Q^2}Q_\mu\right]F_1\left(Q^2\right)
+\frac{m^2_{B}-m^2_P}{Q^2}Q_\mu F_0\left(Q^2\right) ,\nonumber\\
\end{eqnarray}
with $J_{\mu}=\bar{p}_1 \gamma_{\mu}(1-\gamma_5)p_2$, $P_{\mu}=p_{0\mu}+p_{2\mu}$ and $Q_{\mu}=p_{0\mu}-p_{2\mu}$.  $F_{0,1}\left(Q^2\right)$ are the weak transition form factors, and the details would be discussed in the following section.
%while $A_3\left(q^2\right)$ is the linear combination of form factors $A_1\left(q^2\right)$ and $A_2\left(q^2\right)$, which is~\cite{Cheng:2003sm},
%\begin{eqnarray}
%A_3\left(q^2\right)=\frac{m_B+m_{V}}{2m_{V}}A_1\left(q^2\right)-\frac{m_B-m_{V}}{2m_{V}}A_2\left(q^2\right).
%\end{eqnarray}

The amplitudes of the $B^+$ weak transition could be constructed by the products of two hadronic matrix elements. Here, we take the vertex of $B^+ \rightarrow D_{s1}^+\bar{D}^0$ as an example, the amplitude reads,
\begin{eqnarray}
&&\mathcal{A}(B^+ \to D_{s1}^+ \bar{D}^0) =	\mathcal{A}_{\mu}^{B^+ \to D_{s1}^+ \bar{D}^0} \epsilon_{D_{s1}}^\mu  \nonumber\\
&&\qquad = \frac{G_F}{\sqrt{2}}V_{cb}^*V_{cs}a_1	\left \langle D_{s1}^+ \left | J_\mu^\dagger \right |0  \right \rangle 	\left \langle \bar{D}^0 \left | J_\mu \right |B^+  \right \rangle, \hspace{0.5cm}
\end{eqnarray}
where $G_F$ is the Fermi constant, $V_{cb}^*$ and $V_{cs}$ are the Cabibbo-Kobayashi-Maskawa (CKM) matrix elements. $a_1=c_1^{eff}+c_2^{eff}/N_c$ with $c_{1,2}^{eff}$ to be effective Wilson coefficients obtained by the factorization approach\cite{Bauer:1986bm}. In the present estimations, we adopt $G_F=1.166 \times 10^{-5} {\rm GeV}^{-2}$, $V_{cb}=0.041$, $V_{cs}=0.987$ and $a_1=1.05$ as in Refs.~\cite{ParticleDataGroup:2020ssz,Ali:1998eb,Ivanov:2006ni}.

The interaction of light unflavored mesons and heavy-light mesons can be described by effective Lagrangians constructed based on the heavy quark limit and chiral symmetry. The relevant effective Lagrangians are given by,
\begin{eqnarray}
%	&&\mathcal{L}_{\mathcal{D}^*\mathcal{D}^{(*)}\mathcal{P}}=-i g_{\mathcal{D}^{*} \mathcal{D}\mathcal{P}}\left(\mathcal{D}_{i}^{\dagger} \partial_{\mu} \mathcal{P}_{i j} \mathcal{D}_{j}^{* \mu}-\mathcal{D}_{i}^{* \mu \dagger} \partial_{\mu} \mathcal{P}_{i j} \mathcal{D}_{j}\right) \nonumber\\
%	&&\qquad +\frac{1}{2} g_{\mathcal{D}^{*} \mathcal{D}^{*} \mathcal{P}} \varepsilon_{\mu \nu \alpha \beta}\mathcal{D}_{i}^{* \mu \dagger} \partial^{\nu} \mathcal{P}_{i j} \stackrel{\leftrightarrow}{\partial^{\alpha} }\mathcal{D}_{j}^{* \beta}+\text { H.c. },\nonumber\\
	&&\mathcal{L}_{\mathcal{D}^{(*)}\mathcal{D}^{(*)}\mathcal{V}}=-i g_{\mathcal{D} \mathcal{D} \mathcal{V}} \mathcal{D}_{i}^{\dagger}\stackrel{\leftrightarrow}{\partial_{\mu}} \mathcal{D}^{j}\left(\mathcal{V}^{\mu}\right)_{j}^{i} \nonumber\\		&&\qquad -2 f_{\mathcal{D}^{*} \mathcal{D} \mathcal{V}} \varepsilon_{\mu \nu \alpha\beta}(\partial^{\mu}\mathcal{V}^{\nu})_{j}^{i}(\mathcal{D}_{i}^{\dagger} \stackrel{\leftrightarrow}{\partial^{\alpha}} \mathcal{D}^{* \beta j}-\mathcal{D}_{i}^{* \beta \dagger} \stackrel{\leftrightarrow}{\partial^{\alpha}} \mathcal{D}^{j})\nonumber\\
	&&\qquad +i g_{\mathcal{D}^{*} \mathcal{D}^{*} \mathcal{V}} \mathcal{D}_{i}^{* \nu \dagger} \stackrel{\leftrightarrow}{\partial_{\mu}} \mathcal{D}_{v}^{* j}\left(\mathcal{V}^{\mu}\right)_{j}^{i} \nonumber\\
	&&\qquad +4 i f_{\mathcal{D}^{*} \mathcal{D}^{*} \mathcal{V}} \mathcal{D}_{i \mu}^{* \dagger}\left(\partial^{\mu} \mathcal{V}^{\nu}-\partial^{\nu} \mathcal{V}^{\mu}\right)_{j}^{i} \mathcal{D}_{\nu}^{* j}+\text { H.c. },\nonumber\\
%	&&\mathcal{L}_{\mathcal{D}_1^{\prime}\mathcal{D}^{*}\mathcal{P}}=g_{\mathcal{D}_1^{\prime}\mathcal{D}^{*}\mathcal{P}}\Big[\mathcal{D}_{1b\mu}^\prime\stackrel{\leftrightarrow}{\partial^{\mu}}\mathcal{D}_{a\nu}^{*\dagger}\partial^\nu\mathcal{P}_{ba}\nonumber\\
%	&&\qquad -\mathcal{D}_{1b\mu}^\prime\stackrel{\leftrightarrow}{\partial^{\nu}}\mathcal{D}_{a}^{*\mu\dagger}\partial_\nu\mathcal{P}_{ba}+\mathcal{D}_{1b\mu}^\prime \stackrel{\leftrightarrow}{\partial^{\nu}} \mathcal{D}_{\nu}^{*\dagger} \mathcal{P}_{ba} \Big], \nonumber\\
	&&\mathcal{L}_{\mathcal{D}_1^{\prime}\mathcal{D}^{(*)}\mathcal{V}}=g_{\mathcal{D}_1^{\prime}\mathcal{D}\mathcal{V}} \mathcal{D}_b \mathcal{D}_{1\mu a}^{\prime\dagger} \left(\mathcal{V}^\mu\right)_{ba} ,  \nonumber\\
	&&\qquad + i g_{\mathcal{D}_1^{\prime}\mathcal{D}^{*}\mathcal{V}} \varepsilon_{\mu\nu\alpha\beta}\mathcal{D}^{*\mu}\stackrel{\leftrightarrow}{\partial^{\beta}}\mathcal{D}_1^{\prime\alpha\dagger} \left(\mathcal{V}^\nu\right)_{ba}	,  \nonumber\\
	&&\mathcal{L}_{\mathcal{D}_1^{}\mathcal{D}^{(*)}\mathcal{V}}=g_{\mathcal{D}_1^{}\mathcal{D}\mathcal{V}} \mathcal{D}_b \mathcal{D}_{1\mu a}^{\dagger} \left(\mathcal{V}^\mu\right)_{ba} ,  \nonumber\\
	&&\qquad + i g_{\mathcal{D}_1^{}\mathcal{D}^{*}\mathcal{V}} \varepsilon_{\mu\nu\alpha\beta}\mathcal{D}_1^{\nu}\stackrel{\leftrightarrow}{\partial^{\beta}}\mathcal{D}^{*\alpha \dagger} \left(\mathcal{V}^\mu\right)_{ba}	,  \nonumber\\
\end{eqnarray}
where $\mathcal{D}_1^{\prime}=\left( D_1^{\prime}(2430)^0,D_1^{\prime}(2430)^+,D_{s1}^{\prime}(2460)^+ \right)$,$\mathcal{D}_1^{}=\left( D_1(2420)^0,D_1(2420)^+,D_{s1}^{}(2536)^+ \right)$, $\mathcal{D}^{(*)\dagger}=\left(D^{(*)0}, D^{(*)+}, D_s^{(*)+}\right)$  and $A\overset{\leftrightarrow}{\partial}_{\mu}B = A \partial^\mu B - B \partial^\mu A$. $\mathcal{V}$ is the $3\times3$ matrices representation of vector mesons, with their concrete forms being,
\begin{eqnarray}\label{eq:matrix V}
	\mathcal{V} &=& \left(\begin{array}{ccc}\frac{\rho^0} {\sqrt {2}}+\frac {\omega} {\sqrt {2}}&\rho^+ & K^{*+} \\
		\rho^- & -\frac {\rho^0} {\sqrt {2}} + \frac {\omega} {\sqrt {2}} & K^{*0} \\
		K^{*-}& {\bar K}^{*0} & \phi \\
	\end{array}\right).
\end{eqnarray}

\subsection{Decay Amplitude}

With the effective Lagrangians discussed in the previous section, we can now obtain the amplitudes corresponding to the processes shown in Fig.~\ref{fig:hadronlevel}. Here, we present the amplitudes relevant to the $B^+ \to D^+ T_{\bar{c}\bar{s}1}^f(2750)$ process as an example, which read,
\begin{widetext}
\begin{eqnarray}
\mathcal{M}_{D_s^+\bar{D}^0 K^{*0}}&=& i^3 \int\frac{d^4q_3}{(2\pi)^4}\Big[\mathcal{A}^{B^+\rightarrow D_{s}^+ \bar{D}^0}(q_1,q_2)\Big]\Big[-i g_{\mathcal{D}\mathcal{D}\mathcal{V}} (q_1+p_1)_\beta\Big] \Big[i g_{T_{\bar{c}\bar{s}1}D K^*} g_{\phi\theta} \epsilon^\theta(p_2)\Big]\nonumber\\
 &&\times \Big[\frac{1}{q_1^2-m_{q_1}^2}\Big]\Big[\frac{1}{q_2^2-m_{q_2}^2}\Big]  \Big[\frac{-g^{\beta\phi}+q_3^{\beta} q_3^{\phi}/m_{q_3}^2}{q_3^2-m_{q_3}^2}\Big]\mathcal{F}^2(q_3^2,m_{q_3}^2), \nonumber\\
 \mathcal{M}_{D_s^{*+}\bar{D}^0 K^{*0}}&=& i^3 \int\frac{d^4q_3}{(2\pi)^4}\Big[\mathcal{A}_{\mu}^{B^+\rightarrow D_{s}^{*+} \bar{D}^0}(q_1,q_2)\Big]\Big[-2 f_{\mathcal{D}^*\mathcal{D}\mathcal{V}} \mathcal{\epsilon}_{\lambda\beta\kappa \alpha} i q_{3} ^{\lambda}(q_1+p_1)^{\kappa}\Big] \Big[i g_{T_{\bar{c}\bar{s}1}D K^*} g_{\phi\theta} \epsilon^\theta(p_2)\Big]  \nonumber\\
 &&\times\Big[\frac{-g^{\mu\alpha}+q_1^{\mu} q_1^{\alpha}/m_{q_1}^2}{q_1^2-m_{q_1}^2}\Big] \Big[\frac{1}{q_2^2-m_{q_2}^2}\Big]\Big[\frac{-g^{\beta\phi}+q_3^{\beta} q_3^{\phi}/m_{q_3}^2}{q_3^2-m_{q_3}^2}\Big]\mathcal{F}^2(q_3^2,m_{q_3}^2), \nonumber\\
\mathcal{M}_{D_{s1}^{\prime+}\bar{D}^0 K^{*0}} &=& i^3 \int\frac{d^4q_3}{(2\pi)^4}\Big[\mathcal{A}_{\mu}^{B^+\rightarrow D_{s1}^{\prime+} \bar{D}^0}(q_1,q_2)\Big]\Big[i g_{\mathcal{D}_1^\prime \mathcal{D} V }g_{\alpha\beta}\Big] \Big[i g_{T_{\bar{c}\bar{s}1}D K^*} g_{\phi\theta} \epsilon^\theta(p_2)\Big] \nonumber\\
&&\times  \Big[\frac{-g^{\mu\alpha}+q_1^{\mu} q_1^{\alpha}/m_{q_1}^2}{q_1^2-m_{q_1}^2}\Big]\Big[\frac{1}{q_2^2-m_{q_2}^2}\Big] \Big[\frac{-g^{\beta\phi}+q_3^{\beta} q_3^{\phi}/m_{q_3}^2}{q_3^2-m_{q_3}^2}\Big]\mathcal{F}^2(q_3^2,m_{q_3}^2), \nonumber\\
\mathcal{M}_{D_{s1}^{+}\bar{D}^0 K^{*0}}&=& i^3 \int\frac{d^4q_3}{(2\pi)^4}\Big[\mathcal{A}_{\mu}^{B^+\rightarrow D_{s1}^{\prime +} \bar{D}^0}(q_1,q_2)\Big]\Big[i g_{\mathcal{D}_1 \mathcal{D} V}  g_{\alpha\beta}\Big] \Big[i g_{T_{\bar{c}\bar{s}1}D K^*} g_{\phi\theta} \epsilon^\theta(p_2)\Big] \nonumber\\
&&\times \Big[\frac{-g^{\mu\alpha}+q_1^{\mu} q_1^{\alpha}/m_{q_1}^2}{q_1^2-m_{q_1}^2}\Big]\Big[\frac{1}{q_2^2-m_{q_2}^2}\Big]  \Big[\frac{-g^{\beta\phi}+q_3^{\beta} q_3^{\phi}/m_{q_3}^2}{q_3^2-m_{q_3}^2}\Big]\mathcal{F}^2(q_3^2,m_{q_3}^2). 
\label{Eq:amp}
\end{eqnarray}
\end{widetext}

In the above amplitudes, a form factor $\mathcal{F}(q_3^2,m_{q_3}^2)$ in the monopole form is introduced to compensate for the off-shell effect and avoid ultraviolet divergence in loop integrals. Its concrete form is,
\begin{equation}
	\mathcal{F}\left(q_3^2,m_{q_3}^2\right) = \frac{m_{q_3}^2-\Lambda^2}{q_3^2-\Lambda^2}, \label{Eq:FF}
\end{equation}
where the parameter $\Lambda$ can be reparameterized as ${ \Lambda} = m_{q_3} + \alpha {\rm \Lambda_{QCD}}$ with $ \Lambda_{\rm QCD}$ = 220 MeV, and $\alpha$ to be the model parameter, which should be of order unity~\cite{Cheng:2004ru}.

Then, the amplitude of $B^+ \to D^{+} T_{\bar{c}\bar{s}1}^f(2750) $ reads,
\begin{eqnarray}
	\mathcal{M}_{B^+\to  D^{+} T_{\bar{c}\bar{s}1}^f } =\mathcal{M}_{D_{s}^{+}} +\mathcal{M}_{D_{s}^{*+}}+\mathcal{M}_{D_{s1}^{\prime+}}+\mathcal{M}_{D_{s1}^{+}},
\end{eqnarray}
and the partial width of $B^+ \to  D^{+} T_{\bar{c}\bar{s}1}^f(2750)$ can be estimated by,
\begin{eqnarray}
	&&\Gamma_{B^+ \to  D^{+} T_{\bar{c}\bar{s}1}^f } = \frac{1}{8\pi} \frac{|\vec{p}\,|}{m_{B^+}^2}\overline{\left|\mathcal{M}_{B^+\to  D^{+} T_{\bar{c}\bar{s}1}^0 }\right|^2} ,
\end{eqnarray}
where $\vec{p}$ is the momentum of final states in the rest frame of
$B^+$, and the overline above the amplitude indicates the sum over the spins of final states.

\section{Numerical Results and Discussions}
\label{Sec:Num}
\subsection{Coupling Constants}
Before presenting the numerical results, some relevant constants should be clarified. The $f_P$ and $f_V$ in Eq.~\eqref{eq:weak transition} are the decay constants of the charm-strange mesons\cite{FlavourLatticeAveragingGroup:2019iem,Hwang:2004kga,CLEO:1998pnu,Pan:2023hrk}, and their values are taken from different estimates, which are collected in Table.~\ref{tab:decay constants}.

\begin{table}[hbt]
	\caption{The decay constants of the charm-strange mesons considered in the present work.}
	\label{tab:decay constants}
	\renewcommand\arraystretch{1.5}
	\begin{tabular}{p{2.4cm}<\centering p{1.cm}<\centering p{1.cm}<\centering p{1.cm}<\centering p{1.cm}<\centering p{1.cm}<\centering}
		\toprule[1pt]
		%\midrule[0.65pt]
		Decay constants & $f_{D_s}$ & $f_{D_s^*}$ & $f_{D_{s0}}$ & $f_{D_{s1}^\prime}$ & $f_{D_{s1}}$ \\
		\midrule[1pt]
		Value(MeV) & $250$  & $272$ & $70$ & $158$ & $81$  \\
		%\midrule[0.65pt]
		\bottomrule[1pt]	
	\end{tabular}
\end{table}

In addition, the form factors in Eq.~\eqref{eq:weak transition} are usually estimated in the quark model and are known only in the spacelike region. Some methods, such as analytic continuation, could cover the timelike region where the physical decay processes occur. In Refs.~\cite{Cheng:2003sm,Soni:2021fky}, the form factors are parameterized in the form,
\begin{equation}\label{eq:FQzeta}
	F\left(Q^2\right) = \frac{F(0)}{1-a\zeta+b\zeta^2},
\end{equation}
with $\zeta = Q^2/m_{B}^2$. The relevant parameters $F(0)$, $a$, and $b$ for each form factor are collected in Table~\ref{Tab:F0ab}. To avoid ultraviolet divergence in the loop integrals and to evaluate the loop integrals via Feynman parameterization method, we further parameterize these form factors in the form,
\begin{equation}\label{eq:FQlambda}
	F\left(Q^2\right) = F(0)\frac{\Lambda_1^2}{Q^2-\Lambda_1^2}\frac{\Lambda_2^2}{Q^2-\Lambda_2^2}.
\end{equation}
By fitting Eq.~\eqref{eq:FQlambda} with Eq.~\eqref{eq:FQzeta}, we can obtain the values of $\Lambda_1$ and $\Lambda_2$ for each form factor, which are $\Lambda_1=7.75\ \mathrm{GeV},\ \Lambda_2=11.00\ \mathrm{GeV}$ for $F_0$ and $\Lambda_1=6.53\ \mathrm{GeV},\ \Lambda_2=6.84\ \mathrm{GeV}$ for $F_1$, respectively.

\begin{table}
	\caption{\label{Tab:F0ab}Values of parameters $F(0)$, $a$ and $b$ in form factors~\cite{Cheng:2003sm}.}
	\renewcommand\arraystretch{1.5}
	\begin{tabular}{p{1.9cm}<\centering p{1.9cm}<\centering p{1.9cm}<\centering p{1.9cm}<\centering}
		\toprule[1pt]
		%\midrule[0.65pt]
		Parameter &$F(0)$ &$a$ &$b$ \\
		\midrule[1pt]
		$F_0$ &0.67 &0.65 &0.00 \\
		$F_1$ &0.67 &1.25 &0.39 \\

		%\midrule[0.65pt]
		\bottomrule[1pt]	
	\end{tabular}
\end{table}

Considering the chiral symmetry and heavy quark limit, the coupling constants of charmed mesons with vector meson can be related to gauge couplings by~\cite{Isola:2003fh,Falk:1992cx, Casalbuoni:1996pg},
\begin{eqnarray}
	&&g_{\mathcal{D}\mathcal{D}V} = g_{\mathcal{D}^*\mathcal{D}^*V} = \frac{\beta g_V}{\sqrt{2}},~~~f_{\mathcal{D}^*\mathcal{D}V} = \frac{ f_{\mathcal{D}^*\mathcal{D}^*V}}{m_{\mathcal{D}^*}}=\frac{\lambda g_V}{\sqrt{2}},\nonumber\\
	&&g_{\mathcal{D}_0 \mathcal{D}^* V}=-\sqrt{2}g_V \zeta \sqrt{m_{\mathcal{D}} m_{\mathcal{D}_0}},\nonumber\\
	&&g_{\mathcal{D}_1^\prime \mathcal{D} V}=-\sqrt{2}g_V \zeta \sqrt{m_{\mathcal{D}} m_{\mathcal{D}_1^\prime}},~~ g_{\mathcal{D}_1^\prime \mathcal{D}^* V}=\sqrt{2}g_V \zeta,\nonumber\\
	&&g_{\mathcal{D}_1 \mathcal{D} V}=-\frac{2}{\sqrt{3}}g_V \zeta_1 \sqrt{m_{\mathcal{D}} m_{\mathcal{D}_1}},~~ g_{\mathcal{D}_1 \mathcal{D}^* V}=\frac{1}{2\sqrt{3}} g_V \zeta_1,\nonumber\\
\end{eqnarray}
where the parameters are $\beta=0.9$, $\zeta=-\zeta_1=0.1$, $g_V=m_\rho / f_\pi$ with $f_\pi=0.132$ GeV and $h=-0.56$ \cite{Casalbuoni:1996pg}. By matching the
form factor obtained from the light-cone sum rules and lattice QCD, one can obtain the parameter $\lambda = 0.56~{\rm GeV^{-1}}$\cite{Isola:2003fh}.

In Ref.~\cite{Yue:2025ekk}, the authors investigated the decay properties of $T_{\bar{c}\bar{s}1}^f(2750)$ and the coupling constant of $T_{\bar{c}\bar{s}1}^f(2750)$ with its components have been investigated  by taking three typical binding energies, respectively. The estimations indicated that the coupling constants are weakly dependent on the model parameter $\Lambda$, and the values of the coupling constants were estimated to be 5.6 GeV, 7.0 GeV and 8.0 GeV corresponding to $E_b=5,10$ and 15 MeV, respectively~\cite{Yue:2025ekk}.

\subsection{Branching Fractions}

%\begin{figure}[tb]
%	\centering
%	\includegraphics[width=8.5cm]{result-br.pdf}
%	\caption{The branching fraction ratios depending on the model parameter $\alpha$. }
%	\label{Fig:ratios}
%\end{figure}

Now, all the relevant coupling constants and the parameters in the weak transition form factors are prepared except the model parameter $\alpha$ in Eq.~(\ref{Eq:FF}). Since the model parameter $\alpha$ cannot be determined by first-principles methods, it should be of order unity as clarified in Ref.~\cite{Cheng:2004ru}. Generally, the value of this parameter can be determined by comparing the theoretical estimations with the corresponding experimental data. However, the data for the discussed processes are not available yet. Thus, in the present work, we estimate the branching fractions of the involved process by varying the parameter $\alpha$ from 0.5 to 1.5 to discuss the parameter dependences of the predicted branching fractions.

	\begin{figure*}[t]
		\centering
		\includegraphics[width=18.5 cm]{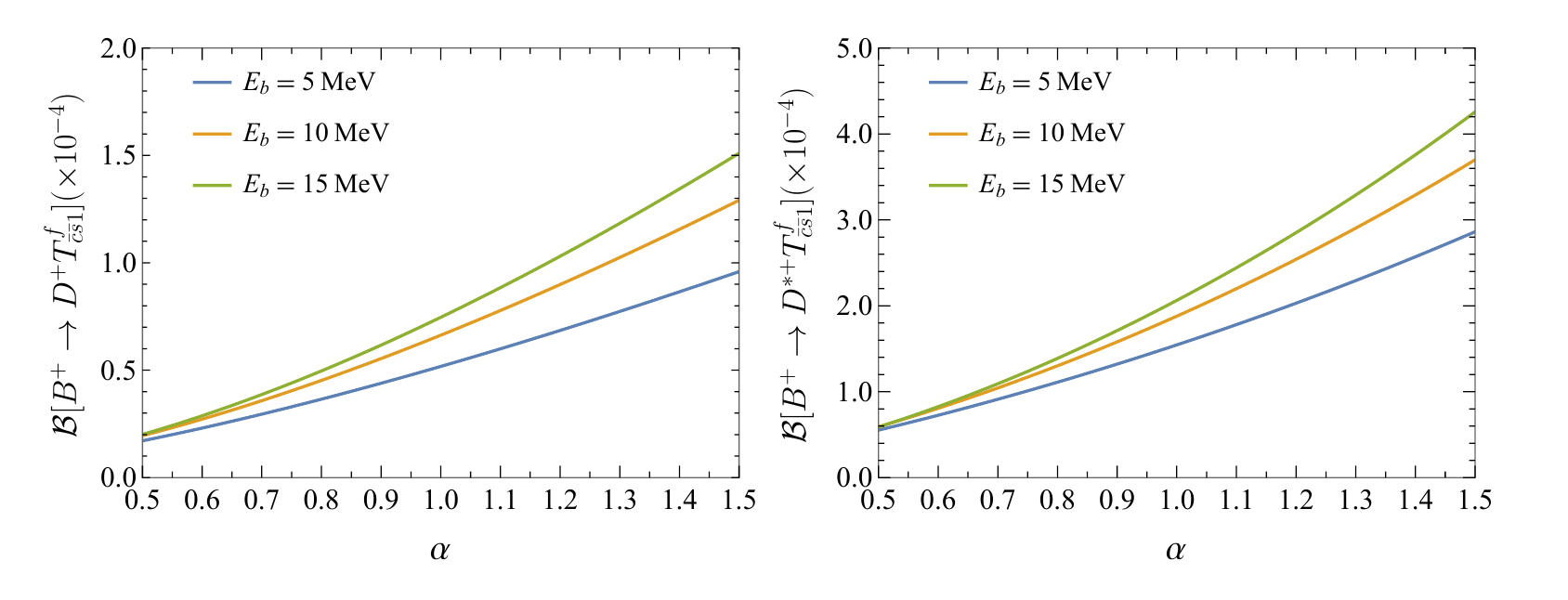}
		\caption{The branching fractions for $B^+ \to D^{+} T_{\bar{c}\bar{s}1}^f(2750) $  (left panel) and $B^+ \to D^{\ast +} T_{\bar{c}\bar{s}1}^f(2750) $ (right panel) depending on the model parameter $\alpha$.}
		\label{Fig:BR}
	\end{figure*}

The $\alpha$ dependence of the branching ratios for $B^+ \to D^{+} T_{\bar{c}\bar{s}1}^f(2750)$ and $B^+ \to D^{\ast+} T_{\bar{c}\bar{s}1}^f(2750) $ are present in Fig.~\ref{Fig:BR} with three typical binding energies, which are $E_b= 5,\ 10,\ 15$ MeV, respectively. From the figure one can find that the branching ratios are estimated to be around $10^{-4}$ and increase with the increasing of parameter $\alpha$ on the whole. In addition, the binding energy  dependences of the branching fractions is found to be mild, In particular, the branching ratios for $B^+ \to D^+ T_{\bar{c}\bar{s}1}^{f}(2750)$ are estimated to be, 
\renewcommand\arraystretch{1.5}
\begin{eqnarray}
&&	\mathcal{B} \left[B^+ \to D^+ T_{\bar{c}\bar{s}1}^f(2750)\right]\nonumber \\
	&&\qquad =\left\{
	\begin{array}{ccl}
	\left(0.52_{-0.35}^{+0.44}\right) \times 10^{-4} && E_b=5\ \mathrm{MeV}\\
	\left(0.66_{-0.47}^{+0.63}\right) \times 10^{-4} && E_b=10\ \mathrm{MeV}\\
	\left(0.75_{-0.55}^{+0.76}\right) \times 10^{-4} && E_b=15\ \mathrm{MeV}\\	
	\end{array}
	\right.,
\end{eqnarray}
while the branching fractions for $B^+ \to D^{\ast +} T_{\bar{c}\bar{s}1}^{f}(2750)$ are evaluated to be,
\renewcommand\arraystretch{1.5}
\begin{eqnarray}
	&&\mathcal{B} \left[B^+ \to D^{\ast + }T_{\bar{c}\bar{s}1}^f(2750)\right]\nonumber\\
	&&\qquad  =\left\{
	\begin{array}{cl}
	\left(1.54_{-0.99}^{+1.32}\right) \times 10^{-4} & E_b=5\ \mathrm{MeV}\\
	\left(1.88_{-1.28}^{+1.82}\right) \times 10^{-4} & E_b=10\ \mathrm{MeV}\\
	\left(2.06_{-1.47}^{+2.19}\right) \times 10^{-4} & E_b=15\ \mathrm{MeV}\\	
	\end{array}
	\right. ,
\end{eqnarray}
where the central values are estimated with $\alpha=1.0$, and the uncertainty comes form the variation of $\alpha$ from 0.5 to 1.5. Our estimations indicate that the branching fractions for $B^+ \to D^{\ast+} T_{\bar{c}\bar{s}1}^f(2750)$ is about three times of that for $B^+ \to D^{+} T_{\bar{c}\bar{s}1}^f(2750)$. In addition, it's worth mentioning that these values are approximately one order of magnitude larger than that of the $B^+\to D^+ T_{\bar{c}\bar{s}1}^{*}(2900)^{0}$ decay~\cite{Yu:2023avh}. 

In Ref.~\cite{Yue:2025ekk}, the authors investigated the decay properties of the $T_{\bar{c}\bar{s}1}^f(2750)$ state in the $\bar{D}K^*$ molecular scenario, and the branching ratio of $T_{\bar{c}\bar{s}1}^f(2750) \to \bar{D}^* K$ was estimated to be $(70.7 ^{+14.0}_{-24.3})\% $, $(68.1 ^{+15.8}_{-26.0})\% $, and $(69.7 ^{+15.9}_{-27.5})\%$ for $E_b=5$, $10$ and $15$ MeV, respectively, which is weakly dependent on the binding energy. Then, the branching fractions of the cascade processes can approximately be,
\begin{eqnarray}
	&&\mathcal{B}[B^+ \to D^{(*)+} T_{\bar{c}\bar{s}1}^f(2750) \to D^{(*)+}\bar{D}^* K ]\nonumber\\
	&&\qquad \simeq \mathcal{B}[B^+ \to   D^{(*)+} T_{\bar{c}\bar{s}1}^f]\times\mathcal{B}[T_{\bar{c}\bar{s}1}^f \to \bar{D}^* K].
\end{eqnarray}
In Table~\ref{Tab:BB}, we present the branching fractions of the above cascade processes.

\begin{table}[t]
	\centering
	\caption{The branching fractions of the cascade process $B^+ \to D^{(*)+} T_{\bar{c}\bar{s}1}^0  \to D^{(*)+}\bar{D}^* K$ estimated in the molecular frame.}
	\label{Tab:BB}
	\renewcommand\arraystretch{1.5}
	\begin{tabular}{p{3.0cm}<\centering p{2.0cm}<\centering p{3.0cm}<\centering}
		\toprule[1pt]
Process & $E_b$ (MeV) & Branching Fraction  \\
		\midrule[1pt]
 & $5$ & $(3.66^{+3.20}_{-2.75}) \times 10^{-5} $ \\
$B^+ \to D^{+}\bar{D}^{*}K$ & $10$ &$(4.51^{+4.41}_{-3.62}) \times 10^{-5}$ \\
	 & $15$ & $(5.20^{+5.46}_{-4.32}) \times 10^{-5}$\\
\midrule[1pt]
%%%%%%	 
 & $5$ & $(1.09^{+0.96}_{-0.80}) \times 10^{-4}$ \\
$B^+ \to D^{*+}\bar{D}^{*}K$ & $10$ & $(1.28^{+1.27}_{-1.00})\times 10^{-4}$ \\
 & $15$ & $(1.44^{+1.56}_{-1.17}) \times 10^{-4}$\\
	\bottomrule[1pt]
	\end{tabular}
\end{table}

%%%%%%%%%%%%%%%%%%%%%%%%%%%%

Furthermore, the considered decay processes $B^+ \to D^{(*)}\bar{D}^*K$ have been measured experimentally with their branching fractions  collected in Table~\ref{tab:fitfraction}. And then, the fit fractions of the $T_{\bar{c}\bar{s}1}^f(2750)$ state in these $B^+$ meson decays processes could be evaluated as,
\begin{eqnarray}\label{eq:FFfunc}
	&&FF[B^+ \to D^{(*)+}T_{\bar{c}\bar{s}1}^f \to D^{(*)+}\bar{D}K] \nonumber\\
	&& \hspace{1cm} =\frac{\mathcal{B}[B^+ \to   D^{(*)+} T_{\bar{c}\bar{s}1}^f]\times\mathcal{B}[T_{\bar{c}\bar{s}1}^f \to \bar{D}^* K]}{\mathcal{B}[B^+ \to D^{(*)+}\bar{D}^*K]}.
\end{eqnarray}
The estimated fit fractions are also presented in Table~\ref{tab:fitfraction}, where the central values of the fit fractions are estimated with $E_b=10$ MeV and $\alpha=1.0$, while the uncertainties are resulted from the variations of the binding energy and the model parameter $\alpha$ as well. From the table, one can find the fit fraction of $T_{\bar{c}\bar{s}1}^f(2750)$ in $B^+ \to D^{+}D^{*-}K^+$ is $\left(3.76_{-3.11}^{+5.19}\right)\%$, which meets expectation of the experimental observation. In addition, the fit fraction of $T_{\bar{c}\bar{s}1}^f(2750)$ in  $B^+ \to D^{*+}D^{*-}K^+$ process is $\left(4.85_{-3.81}^{+6.55}\right)\%$. In this way, we can conclude that if the structure observed by LHCb Collaboration in $B^+ \to D^{+}D^{*-}K^+$ process was $T_{\bar{c}\bar{s}1}^f(2750)$ state, it could also be observed in $B^+ \to D^{*+}D^{*-}K^+$ process.

\begin{table}[t]
	\centering
	\caption{The branching ratios of $B^+ \to D^{(*)+}\bar{D}^*K$ in PDG average and the fit fraction estimated in the presented work.}
	\label{tab:fitfraction}
	\renewcommand\arraystretch{1.5}
	\begin{tabular}{p{3.0cm}<\centering p{2.5cm}<\centering p{2.5cm}<\centering}
		\toprule[1pt]
		Process & Branching ratio & FF(\%) \\
		\midrule[1pt]
		$B^+ \to D^{+}D^{*-}K^+$ & $(6.0\pm 1.3)\times 10^{-4}$ & $3.76_{-3.11}^{+5.19}$ \\
		$B^+ \to D^{+}\bar{D}^{*0}K^0$ & $(2.1\pm 0.5)\times 10^{-3}$ & $1.07_{-0.89}^{+1.49}$\\
		$B^+ \to D^{*+}D^{*-}K^+$ & $(1.32\pm 0.18 )\times 10^{-3}$ & $4.85_{-3.81}^{+6.55}$\\
		$B^+ \to D^{*+}\bar{D}^{*0}K^0$ & $(9.2\pm 1.2)\times 10^{-3}$ & $0.70_{-0.55}^{+0.94}$ \\
		\bottomrule[1pt]
	\end{tabular}
\end{table}

%\begin{table}[tb]
%	\centering
%	\caption{The fit fraction of $B^+ \to D^{(*)+}\bar{D}^*K$}
%	\label{tab:ff}
%	\begin{tabular}{p{4.0cm}<\centering p{2.5cm}<\raggedright}
%		\toprule[1pt]
%		Process & Branching ratio \\
%		\midrule[1pt]
%		$FF[B^+ \to  D^{+}T_{\bar{c}\bar{s}1}^0 \to D^{+}D^{*-}K^+]$ & $(6.0\pm 1.3)\times 10^{-4}$ \\
%		$FF[B^+ \to  D^{+}T_{\bar{c}\bar{s}1}^0 \to D^{+}\bar{D}^{*0}K^0]$ & $(2.1\pm 0.5)\times 10^{-3}$ \\
%		$FF[B^+ \to  D^{*+}T_{\bar{c}\bar{s}1}^0 \to D^{*+}D^{*-}K^+]$ & $(1.32\pm 0.18)\times 10^{-3}$ \\
%		$FF[B^+ \to  D^{*+}T_{\bar{c}\bar{s}1}^0\to D^{*+}\bar{D}^{*0}K^0]$ & $(9.2\pm 1.2)\times 10^{-3}$ \\
%		\bottomrule[1pt]
%	\end{tabular}
%\end{table}

\section{SUMMARY}
\label{Sec:sum}
In the molecular state scenario, the states near the $D^{(*)}K^{(*)}/\bar{D}^{(*)}K^{(*)}$ threshold have attracted much attention. Predictions of molecular states near the $DK,D^*K$ and $D^*K^*$ thresholds have been experimentally observed, while the $DK^*$ molecular state have also been predicted but has not been observed yet. In the present work, we assumed $T_{\bar{c}\bar{s}1}^f(2750)$ as a $DK^*$ molecular state with three typical binding energy $E_b=5,10$ and 15 MeV, and investigated its production in the $B^+$ decay process through the meson loop mechanism. The branching ratio of $B^+ \to T_{\bar{c}\bar{s}1}^f(2750) D^{+}$ was estimated to be $(0.66_{-0.47}^{+0.63}) \times 10^{-4}$ with $E_b=10$ MeV, which are significantly larger than that of $B^+\to D^+ T_{\bar{c}\bar{s}1}^{*}(2900)^{0}$ by about an order of magnitude. 

Besides $B^+ \to T_{\bar{c}\bar{s}1}^f(2750) D^{+}$, we also considered the case with a charmed meson $D^{*+}$ in final state, and the branching ratio of $B^+ \to T_{\bar{c}\bar{s}1}^f(2750) D^{*+}$ is  $(1.88_{-1.28}^{+1.82}) \times 10^{-4}$ with $E_b=10$ MeV. Using the measured branching ratio of $B^+ \to D^{(*)+}D^{*-}K^+$ and $B^+ \to D^{(*)+}\bar{D}^{0}K^0$ from the PDG, we estimated the fit fraction of $T_{\bar{c}\bar{s}1}^f(2750)$ in $B^+ \to D^{+}D^{*-}K^+$ to be $3.76_{-3.11}^{+5.19}\%$ in the $\bar{D}K^*$ molecular scenario. The fit fraction of $T_{\bar{c}\bar{s}1}^f(2750)$ in  $B^+ \to D^{*+}D^{*-}K^+$ is estimated to be   $4.85_{-3.81}^{+6.55}\%$, which is comparable to that in the $B^+ \to D^{+}D^{*-}K^+$ process. Thus, we suggest searching for $T_{\bar{c}\bar{s}1}^f(2750)$ in $B^+ \to D^{*+}D^{*-}K^+$.

In summary, we stress that our numerical evaluations are based on the molecular ansatz for $T_{\bar{c}\bar{s}1}^{f}(2750)$, whereby this state is interpreted as a $\bar{D}K^*$ molecular state and all relevant coupling constants are determined within the molecular picture. Accordingly, our results cannot preclude compact-tetraquark interpretations nor the interpretation via the triangle-singularity mechanism. Further systematic studies of the relevant processes incorporating compact-tetraquark scenarios and triangle-singularity dynamics will shed light on the intrinsic nature of the candidate $T_{\bar{c}\bar{s}1}^{f}(2750)$ state.

\section*{ACKNOWLEDGMENTS}
 This work is supported by the National Natural Science Foundation of China under the Grant Nos. 12175037, 12335001 and 12405093, as well as supported, in part, by National Key Research and Development Program under Grant Nos. No.2024YFA1610503 and 2024YFA1610504.

\bibliographystyle{unsrt}
\bibliography{references}
%\nocite{*}

\end{CJK}	
\end{document}